# Effect of both Z and $Z'$-mediated flavor-changing neutral currents on the baryonic rare decay $\Lambda_b \to \Lambda \ell^+ \ell^-$

**S Sahoo[1], C K Das[2] and L Maharana[3]**

[1]Department of Physics, National Institute of Technology,
Durgapur-713 209, West Bengal, India
E-mail: sukadevsahoo@yahoo.com

[2]Department of Physics, Trident Academy of Technology,
Bhubaneswar – 751 024, Orissa, India.
E-mail: c.das13@rediffmail.com

[3]Department of Physics, Utkal University,
Bhubaneswar – 751 004, Orissa, India.
E-mail: lmaharan@iopb.res.in

**Abstract**

We study the effect of both Z and $Z'$-mediated flavor-changing neutral currents (FCNCs) on the $\Lambda_b \to \Lambda \ell^+ \ell^-$ $(\ell = \mu, \tau)$ rare decay. We find the branching ratio is reasonably enhanced from its standard model value due to the effect of both Z and $Z'$-mediated FCNCs, and gives the possibility of new physics beyond the standard model. The contribution of $Z'$-boson depends upon the precise value of $M_{Z'}$.
.

## 1. Introduction

Rare B decays [1,2] induced by flavor-changing neutral current (FCNC) $b \to s(d)\ell^+\ell^-$ transitions are very important to probe the flavor sector of the standard model (SM). In the SM, they arise from one-loop diagrams and are generally suppressed in comparison to the tree diagrams. Nevertheless, one-loop FCNC processes can be enhanced by orders of magnitude in some cases due to the presence of new physics. New physics (NP) comes into play in rare B decays in two different ways: (a) through a new contribution to the Wilson coefficients or (b) through a new structure in the effective Hamiltonian, which are both absent in the SM. Rare decays can give valuable information about the Cabibbo-Kobayashi-Maskawa (CKM) matrix elements, $V_{td}, V_{ts}, V_{tb}$, etc and leptonic decay constants. Moreover, $b \to s(d)\ell^+\ell^-$ decay is very sensitive to the new physics beyond the SM.

The study of $b \to s\ell^+\ell^-$ decays is one of the most reliable tests of FCNC. These decays have been studied in the SM, two-Higgs-doublet model (2HDM) and minimal supersymmetric standard model (MSSM) [3-14]. The theoretical study of the inclusive decays is easy but their experimental detection is quite difficult. For exclusive decays the situation is opposite i.e. their experimental detection is easy, but theoretical analysis is very difficult. One of the exclusive decay which is described at inclusive level by $b \to s\ell^+\ell^-$ transition is the baryonic $\Lambda_b \to \Lambda\ell^+\ell^-$ ($\ell = e, \mu, \tau$) decay. This decay has been studied in the SM [15], in the supersymmetric model with and without R-parity [16-19], in the two-Higgs-doublet model [20] and in a model independent way [21]. In comparison with B meson decays, $\Lambda_b$ baryon decays contain some particular observables, involving the spin of the b quark, which are sensitive to new physics and more easily detectable [22].

Theoretically it is predicted that $Z'$-bosons are exist in grand unified theories (GUTs), superstring theories and theories with large extra dimensions but experimentally the $Z'$-boson is not discovered so far [23]. If the $Z'$-bosons couple to quarks and leptons not too weakly and if their mass is not too large, they will be produced at the Tevatron and the LHC and easily detected through their leptonic decay modes [24]. Therefore the search for these particles is a very challenging topic in experimental physics.

There has been rigorous study in the $Z'$ sector to understand physics beyond the SM [25-28]. It has been shown that a leptophobic $Z'$-boson can appear in $E_6$ gauge models due to mixing of gauge kinetic terms [29,30]. Flavor mixing can be induced at the tree level in the up-type and/or down-type quark sector after diagonalizing their mass matrices. Mixing between ordinary and exotic left-handed quarks induces Z-mediated FCNCs. The right-handed quarks $d_R, s_R$ and $b_R$ have different $U(1)'$ quantum numbers than exotic $q_R$ and their mixing will induce $Z'$-mediated FCNCs [29,31,32] among the ordinary down quark types. Tree level FCNC interactions can also be induced by an additional $Z'$-boson on the up-type quark sector [33]. In the $Z'$ model [34], the FCNC $b-s-Z'$ coupling is related to the flavor-diagonal couplings $qqZ'$ in a predictive way, which is then used to obtain upper limits on the leptonic $\ell\ell Z'$ couplings. Hence, it is possible to predict the branching ratio for $\Lambda_b \to \Lambda \ell^+ \ell^-$ rare decay. With FCNCs, both Z and $Z'$-boson contributes at tree level, and its contribution will interfere with the SM contributions [31,32,35]. In this paper, we study $\Lambda_b \to \Lambda \ell^+ \ell^-$ rare decay considering the effect of both Z and $Z'$-mediated FCNCs that change the effective Hamiltonian and modifies the branching ratio.

This paper is organized as follows: in Section 2, we discuss the $\Lambda_b \to \Lambda \ell^+ \ell^-$ rare decay in the standard model. In Section 3, we give a brief account of the extended quark sector model and explain why it implies FCNC at the tree level. Then, we evaluate the effective Hamiltonian for $\Lambda_b \to \Lambda \ell^+ \ell^-$ rare decay considering the contributions from both the Z and $Z'$-bosons. In Section 4, we calculate the branching ratio for $\Lambda_b \to \Lambda \ell^+ \ell^-$ decay. Then we discuss our results, so obtained.

## 2. $\Lambda_b \to \Lambda \ell^+ \ell^-$ Decay in the Standard Model

Let us consider the $\Lambda_b \to \Lambda \ell^+ \ell^-$ $(\ell = \mu, \tau)$ rare decay process. In the standard model, this process is loop-suppressed. However, it is potentially sensitive to new physics beyond the SM. At the quark level, the $\Lambda_b \to \Lambda \ell^+ \ell^-$ decay is described by the $b \to s \ell^+ \ell^-$ transition. The matrix element of the $b \to s \ell^+ \ell^-$ process contains terms describing the virtual effects induced by $t\bar{t}, c\bar{c}$ and $u\bar{u}$ loops which are proportional to $V_{tb}V_{ts}^*, V_{cb}V_{cs}^*$ and $V_{ub}V_{us}^*$ respectively. From unitarity of the CKM matrix and neglecting $V_{ub}V_{us}^*$ in comparison to $V_{tb}V_{ts}^*$ and $V_{cb}V_{cs}^*$, it is clear that the matrix element of the $b \to s \ell^+ \ell^-$ contains only one independent CKM factor $V_{tb}V_{ts}^*$. The effective Hamiltonian describing $\Lambda_b \to \Lambda \ell^+ \ell^-$ decay process is given [2,36]:

$$H_{eff} = \frac{G_F \alpha}{\sqrt{2} \pi} \lambda_t [ C_9^{eff} (\bar{s} \gamma_\mu P_L b)(\bar{\ell} \gamma^\mu \ell) + C_{10} (\bar{s} \gamma_\mu P_L b)(\bar{\ell} \gamma^\mu \gamma_5 \ell) \\ - 2 C_7^{eff} m_b \left( \bar{s} i \sigma_{\mu\nu} \frac{p^\mu}{p^2} P_R b \right)(\bar{\ell} \gamma^\mu \ell) ],$$

(1)

where $G_F$ is the Fermi coupling constant, $\alpha$ is the electromagnetic coupling constant, $\lambda_t = V_{tb}V_{ts}^*$, $P_{R,L} = \frac{1}{2}(1 \pm \gamma_5)$, $p$ is the momentum transferred to the lepton pair and $p = p_+ + p_-$ the sum of the momenta of the $\ell^+$ and $\ell^-$, and $C_7^{eff}$, $C_9$ and $C_{10}$ are Wilson coefficients [37] evaluated at the b quark mass scale in the modified minimal subtraction $(\overline{MS})$ scheme ($m_b = 4.6$ GeV).

$$C_7^{eff} = -0.308, \quad C_9 = 4.154, \quad C_{10} = -4.261.$$

(2)

The coefficient $C_9^{eff}$ has a perturbative part and a resonance part which comes from the long distance effects. Therefore, we can write

$$C_9^{eff} = C_9 + Y(s) + C_9^{res}, \qquad (3)$$

where $s = p^2$ and the function $Y(s)$ is the perturbative part coming from one loop matrix elements of the four quark operators and is given by [20,38]

$$Y(s) = g(m_c, s)(3C_1 + C_2 + 3C_3 + C_4 + 3C_5 + C_6) - \frac{1}{2} g(0, s)(C_3 + 3C_4)$$
$$- \frac{1}{2} g(m_b, s)(4C_3 + 4C_4 + 3C_5 + C_6) + \frac{2}{9}(3C_3 + C_4 + 3C_5 + C_6), \qquad (4)$$

where

$$g(m_j, s) = -\frac{8}{9} \ln(m_j / m_b^{pole}) + \frac{8}{27} + \frac{4}{9} y_j - \frac{2}{9}(2 + y_j)\sqrt{|1 - y_j|} \times$$
$$\left[ \theta(1 - y_j) \left\{ \ln\left(\frac{1 + \sqrt{1 - y_j}}{1 - \sqrt{1 - y_j}}\right) - i\pi \right\} + \theta(y_j - 1) 2 \arctan \frac{1}{\sqrt{y_j - 1}} \right], \qquad (5)$$

with $y_j = 4m_j^2 / s$. The values of the coefficients $C_i$ in next-to-leading-logarithmic (NLL) order are taken from [37] as $C_1 = -0.151$, $C_2 = 1.059$, $C_3 = 0.012$, $C_4 = -0.034$, $C_5 = 0.010$ and $C_6 = -0.040$. The long distance resonance effect is given as [5,8,39]

$$C_9^{res} = \frac{3\pi}{\alpha^2}(3C_1 + C_2 + 3C_3 + C_4 + 3C_5 + C_6) \times \sum_{v_i = \psi(1S),\ldots,\psi(6S)} k v_i \frac{m_{v_i} \Gamma(V_i \to \ell^+ \ell^-)}{m_{v_i}^2 - s - i m v_i \Gamma v_i}, \qquad (6)$$

where the phenomenological parameter $k$ [9] is satisfied the relation $k(3C_1 + C_2) = -1$.

Now the amplitude of $\Lambda_b(p_{\Lambda_b}) \to \Lambda(p_\Lambda) \ell^+(p_+) \ell^-(p_-)$ decay can be obtained by sandwiching $H_{eff}$ for the $b \to s \ell^+ \ell^-$ transition between initial and final baryon states i.e. $\langle \Lambda | H_{eff} | \Lambda_b \rangle$. The matrix elements of the various hadronic currents between the initial $\Lambda_b$ and the final $\Lambda$ baryon can be written as [2]:

$$\langle \Lambda | \bar{s} \gamma_\mu b | \Lambda_b \rangle = \bar{u}_\Lambda \left[ f_1 \gamma_\mu + i f_2 \sigma_{\mu\nu} p^\nu + f_3 p_\mu \right] u_{\Lambda_b},$$

$$\langle \Lambda | \bar{s} \gamma_\mu \gamma_5 b | \Lambda_b \rangle = \bar{u}_\Lambda \left[ g_1 \gamma_\mu \gamma_5 + i g_2 \sigma_{\mu\nu} \gamma_5 p^\nu + g_3 \gamma_5 p_\mu \right] u_{\Lambda_b},$$

$$\langle \Lambda | \bar{s} i \sigma_{\mu\nu} p^\nu b | \Lambda_b \rangle = \bar{u}_\Lambda \left[ f_1^T \gamma_\mu + i f_2^T \sigma_{\mu\nu} p^\nu + f_3^T p_\mu \right] u_{\Lambda_b},$$

$$\langle \Lambda | \bar{s} i \sigma_{\mu\nu} \gamma_5 p^\nu b | \Lambda_b \rangle = \bar{u}_\Lambda \left[ g_1^T \gamma_\mu \gamma_5 + i g_2^T \sigma_{\mu\nu} \gamma_5 p^\nu + g_3^T \gamma_5 p_\mu \right] u_{\Lambda_b}, \qquad (7)$$

where $p = p_{\Lambda_b} - p_\Lambda = p_+ + p_-$ is the momentum transfer, and $f_i$ and $g_i$ are the various form factors which are functions of $p^2$. The processes for the heavy to light baryonic decays such as those with $\Lambda_b \to \Lambda$ have been studied based on the heavy quark effective theory (HQET) in [40]. It is found that the number of independent form factors is reduced to two in the heavy quark symmetry limit. In this limit, the matrix elements of all hadronic currents, irrespective of their Dirac structure, can be written as

$$\langle \Lambda(p_\Lambda) | \bar{s} \Gamma b | \Lambda_b(p_{\Lambda_b}) \rangle = \bar{u}_\Lambda [F_1(p^2) + \slashed{v} F_2(p^2)] \Gamma u_{\Lambda_b}, \tag{8}$$

where $\Gamma$ is the product of Dirac matrices, $v^\mu = p^\mu_{\Lambda_b}/m_{\Lambda_b}$ is the four velocity of $\Lambda_b$ and $F_{1,2}$ are the form factors. The relations among these two sets of form factors are given as [16-19]:

$$g_1 = f_1 = f_2^T = g_2^T = F_1 + \sqrt{r} F_2, \qquad g_2 = f_2 = g_3 = f_3 = \frac{F_2}{m_{\Lambda_b}},$$

$$g_3^T = \frac{F_2}{m_{\Lambda_b}}(m_{\Lambda_b} + m_\Lambda), \quad f_3^T = -\frac{F_2}{m_{\Lambda_b}}(m_{\Lambda_b} - m_\Lambda), \quad f_1^T = g_1^T = \frac{F_2}{m_{\Lambda_b}} p^2, \tag{9}$$

where $r = m_\Lambda^2 / m_{\Lambda_b}^2$. The form factors $F_1$ and $F_2$ for the $\Lambda_b \to \Lambda \ell^+ \ell^-$ decay are calculated in QCD sum rule approach combined with heavy quark symmetry in [15,16] and values are also presented in [21]. The values of form factors are $F_1 = 0.462$ and $F_2 = -0.077$. Using the recent data [16,41] we get the ratio of form factors, $r = -0.20 \pm 0.04$. From equation (9), it is clear that $f_1(g_1)$ and $f_2^T(g_2^T)$ are proportional to $F_1$. Hence, they are large whereas all others are small. Now using these form factors, the transition amplitude can be written as [2]:

$$M(\Lambda_b \to \Lambda \ell^+ \ell^-) = \frac{G_F \alpha}{\sqrt{2} \pi} \lambda_t \times \begin{bmatrix} \bar{\ell} \gamma_\mu \ell \{ \bar{u}_\Lambda [\gamma^\mu (A_1 P_R + B_1 P_L) + i\sigma^{\mu\nu} p_\nu (A_2 P_R + B_2 P_L)] u_{\Lambda_b} \} \\ + \bar{\ell} \gamma_\mu \gamma_5 \ell \left\{ \bar{u}_\Lambda \begin{bmatrix} \gamma^\mu (D_1 P_R + E_1 P_L) + i\sigma^{\mu\nu} p_\nu (D_2 P_R + E_2 P_L) \\ + p^\mu (D_3 P_R + E_3 P_L) \end{bmatrix} u_{\Lambda_b} \right\} \end{bmatrix}$$

............(10)

where the parameters $A_i$, $B_i$ and $D_j$, $E_j$ ($i = 1, 2$ and $j = 1, 2, 3$) are defined as

$$A_i = \frac{1}{2} C_9^{\text{eff}} (f_i - g_i) - \frac{2 C_7^{\text{eff}} m_b}{p^2} (f_i^T + g_i^T),$$

$$B_i = \frac{1}{2} C_9^{\text{eff}} (f_i + g_i) - \frac{2 C_7^{\text{eff}} m_b}{p^2} (f_i^T - g_i^T),$$

$$D_j = \frac{1}{2} C_{10} (f_j - g_j), \qquad E_j = \frac{1}{2} C_{10} (f_j + g_j). \tag{11}$$

From equation (11), it is clear that $f^T$ and $g^T$ are associated with $C_7$ which is about one order magnitude smaller than $C_9$ and $C_{10}$ [equation (2)] so that their effects due to the deviation of the results in the HQET are small [18].

**Differential Decay Rates and Branching Ratios:**

The double partial decay rates for $\Lambda_b \to \Lambda \ell^+ \ell^-$ $(\ell = \mu, \tau)$ can be obtained from the transition amplitude [equation (10)] as:

$$\frac{d^2\Gamma}{d\hat{s}\,dz} = \frac{G_F^2 \alpha^2 \lambda_t^2}{2^{12}\pi^5} m_{\Lambda_b} \sqrt{1 - \frac{4m_l^2}{p^2}} \sqrt{\lambda(1, r, \hat{s})}\, K(s, z)\,, \tag{12}$$

where $\hat{s} = s/m_{\Lambda_b}^2$, $z = \cos\theta$, the angle between $p_{\Lambda_b}$ and $p_+$ in the center of mass frame of $\ell^+ \ell^-$ pair, and $\lambda(a, b, c) = a^2 + b^2 + c^2 - 2(ab + bc + ca)$ is the usual triangle function. The function $K(s, z)$ is given as

$$K(s, z) = K_0(s) + z\, K_1(s) + z^2\, K_2(s)\,, \tag{13}$$

where

$$\begin{aligned}
K_0(s) &= 32 m_l^2 m_{\Lambda_b}^4 \hat{s}(1 + r - \hat{s})\left(|D_3|^2 + |E_3|^2\right) + 64 m_l^2 m_{\Lambda_b}^3 (1 - r - \hat{s})\mathrm{Re}\left(D_1^* E_3 + D_3 E_1^*\right) \\
&\quad + 64 m_{\Lambda_b}^2 \sqrt{r}\left(6 m_l^2 - \hat{s}\, m_{\Lambda_b}^2\right)\mathrm{Re}\left(D_1^* E_1\right) \\
&\quad + 64 m_l^2 m_{\Lambda_b}^3 \sqrt{r} \times \left[2 m_{\Lambda_b} \hat{s}\, \mathrm{Re}\left(D_3^* E_3\right) + (1 - r + \hat{s})\mathrm{Re}\left(D_1^* D_3 + E_1^* E_3\right)\right] \\
&\quad + \\
&\quad 32 m_{\Lambda_b}^2 \left(2 m_l^2 + \hat{s}\, m_{\Lambda_b}^2\right) \times \begin{bmatrix} (1 - r + \hat{s})\, m_{\Lambda_b} \sqrt{r}\, \mathrm{Re}\left(A_1^* A_2 + B_1^* B_2\right) - m_{\Lambda_b}(1 - r - \hat{s})\mathrm{Re}\left(A_1^* B_2 + A_2^* B_1\right) \\ -2\sqrt{r}\left\{\mathrm{Re}\left(A_1^* B_1\right) + m_{\Lambda_b}^2 \hat{s}\, \mathrm{Re}\left(A_2^* B_2\right)\right\} \end{bmatrix} \\
&\quad + 8 m_{\Lambda_b}^2\left[4 m_l^2 (1 + r - \hat{s}) + m_{\Lambda_b}^2 \left\{(1 - r)^2 - \hat{s}^2\right\}\right] \times \left(|A_1|^2 + |B_1|^2\right) \\
&\quad + 8 m_{\Lambda_b}^4\left[4 m_l^2 \left\{\lambda + (1 + r - \hat{s})\hat{s}\right\} + m_{\Lambda_b}^2 \hat{s}\left\{(1 - r)^2 - \hat{s}^2\right\}\right] \times \left(|A_2|^2 + |B_2|^2\right) \\
&\quad - 8 m_{\Lambda_b}^2\left[4 m_l^2 (1 + r - \hat{s}) - m_{\Lambda_b}^2 \left\{(1 - r)^2 - \hat{s}^2\right\}\right] \times \left(|D_1|^2 + |E_1|^2\right) \\
&\quad + 8 m_{\Lambda_b}^5 \hat{s}\, v_l^2 \times \\
&\quad \begin{bmatrix} -8 m_{\Lambda_b} \hat{s}\sqrt{r}\,\mathrm{Re}\left(D_2^* E_2\right) + 4(1 - r + \hat{s})\sqrt{r}\,\mathrm{Re}\left(D_1^* D_2 + E_1^* E_2\right) \\ -4(1 - r - \hat{s})\mathrm{Re}\left(D_1^* E_2 + D_2^* E_1\right) + m_{\Lambda_b}\left\{(1 - r)^2 - \hat{s}^2\right\}\left\{|D_2|^2 + |E_2|^2\right\} \end{bmatrix},
\end{aligned} \tag{14}$$

$$K_1(s) = -16 m_{\Lambda_b}^4 \hat{s}\, v_l \sqrt{\lambda}\left\{2\,\mathrm{Re}\left(A_1^* D_1\right) - 2\,\mathrm{Re}\left(B_1^* E_1\right) + 2 m_{\Lambda_b}\,\mathrm{Re}\left(B_1^* D_2 - B_2^* D_1 + A_2^* E_1 - A_1^* E_2\right)\right\}$$

$$+ 32 m_{\Lambda_b}^5 \hat{s} v_l \sqrt{\lambda} \left\{ m_{\Lambda_b} (1-r) \text{Re}(A_2^* D_2 - B_2^* E_2) + \sqrt{r} \text{Re}(A_2^* D_1 + A_1^* D_2 - B_2^* E_1 - B_1^* E_2) \right\}$$
......(15)

and

$$K_2(s) = 8 m_{\Lambda_b}^6 \hat{s} v_l^2 \lambda \left[ |A_2|^2 + |B_2|^2 + |D_2|^2 + |E_2|^2 \right]$$
$$- 8 m_{\Lambda_b}^4 v_l^2 \lambda \left[ |A_1|^2 + |B_1|^2 + |D_1|^2 + |E_1|^2 \right], \tag{16}$$

where $\lambda$ is the short form for $\lambda(1, r, \hat{s})$. Now integrating equation (12) w. r. t. the angular dependent parameter z, we can get the expression for decay width as :

$$\left( \frac{d\Gamma}{ds} \right)_0 = \frac{G_F^2 \alpha^2 \lambda_t^2}{2^{11} \pi^5 m_{\Lambda_b}} \sqrt{1 - \frac{4 m_l^2}{p^2}} \sqrt{\lambda(1, r, \hat{s})} \left[ K_0(s) + \frac{1}{3} K_2(s) \right], \tag{17}$$

The limits for *s* are based on the kinematic phase space:

$$4 m_l^2 \leq s \leq (m_{\Lambda_b} - m_\Lambda)^2. \tag{18}$$

The branching ratios can be calculated by multiplying decay width $\Gamma$ with the life time of $\Lambda_b$ i.e.

$$B(\Lambda_b \to \Lambda \ell^+ \ell^-) = \Gamma(\Lambda_b \to \Lambda \ell^+ \ell^-) \tau_{\Lambda_b}, \tag{19}$$

where $\tau_{\Lambda_b}$ is the life time of $\Lambda_b$ and $\tau_{\Lambda_b} = \left( 1.383^{+0.049}_{-0.048} \right) \times 10^{-12}$ s [41]. The value of the branching ratios for $\Lambda_b \to \Lambda \ell^+ \ell^-$ decay in the standard model [2] is given as:

$$B(\Lambda_b \to \Lambda \mu^+ \mu^-) = 4.55 \times 10^{-6},$$
$$B(\Lambda_b \to \Lambda \tau^+ \tau^-) = 0.17 \times 10^{-6}. \tag{20}$$

These branching ratios have also been calculated in an explicit supersymmetric standard model [18] as:

$$B(\Lambda_b \to \Lambda \mu^+ \mu^-) = 2.24 \times 10^{-6},$$
$$B(\Lambda_b \to \Lambda \tau^+ \tau^-) = 0.179 \times 10^{-6}. \tag{21}$$

### 3. The Model

In extended quark sector model [35], besides the three standard generations of the quarks, there is an $SU(2)_L$ singlet of charge $-1/3$. This model allows for Z-mediated FCNCs. The up quark sector interaction eigenstates are identified with mass eigenstates but down quark sector interaction eigenstates are related to the mass eigenstates by a 4 × 4 unitary matrix, which is denoted by K. The charged-current interactions are described by

$$L_{\text{int}}^W = \frac{g}{\sqrt{2}} \left( W_\mu^- J^{\mu^+} + W_\mu^+ J^{\mu^-} \right), \tag{22}$$

$$J^{\mu^-} = V_{ij}\bar{u}_{iL}\gamma^\mu d_{jL}. \tag{23}$$

The charged-current mixing matrix V is a 3 × 4 submatrix of K :

$$V_{ij} = K_{ij} \quad \text{for } i = 1,\ldots 3, \quad j = 1,\ldots,4. \tag{24}$$

Here, V is parametrized by six real angles and three phases, instead of three angles and one phase in the original CKM matrix.

The neutral-current interactions are described by

$$L_{int}^Z = \frac{g}{\cos\theta_W} Z_\mu \left(J^{\mu 3} - \sin^2\theta_W J_{em}^\mu\right), \tag{25}$$

$$J^{\mu 3} = -\frac{1}{2} U_{pq} \bar{d}_{pL}\gamma^\mu d_{qL} + \frac{1}{2}\delta_{ij}\bar{u}_{iL}\gamma^\mu u_{jL}. \tag{26}$$

In neutral-current mixing, the matrix for the down sector is $U = V^\dagger V$. Since in this case V is not unitary, $U \neq 1$. Its nondiagonal elements do not vanish:

$$U_{pq} = - K_{4p}^* K_{4q} \quad \text{for } p \neq q. \tag{27}$$

Since the various $U_{pq}$ are non-vanishing, they allow for flavor-changing neutral currents that would be a signal for new physics.

Now consider the $\Lambda_b \to \Lambda \ell^+ \ell^-$ decay process in the presence of Z-mediated FCNC [35] at tree level. The $Zbs$ FCNC coupling, which affects B-decays, is parameterized by one independent parameter $U_{sb}$ and this parameter is constrained by branching ratio of the process $B \to X_S \ell^+ \ell^-$ and is found to be $|U_{sb}| \cong 1\times 10^{-3}$ [42]. The BELLE Collaboration [43] has measured the branching ratio $B(B \to X_S \ell^+ \ell^-) = (6.1 \pm 1.4^{+1.4}_{-1.1}) \times 10^{-6}$. Considering the contribution of the Z-boson to $\Lambda_b \to \Lambda \ell^+ \ell^-$ ($\ell = \mu, \tau$) decay, one can write the effective Hamiltonian [2] as

$$H_{eff}(Z) = \frac{G_F}{\sqrt{2}} U_{sb} \left[\bar{s}\gamma^\mu(1-\gamma_5)b\right]\left[\bar{\ell}\left(C_V^\ell \gamma_\mu - C_A^\ell \gamma_\mu\gamma_5\right)\ell\right], \tag{28}$$

where $C_V^\ell$ and $C_A^\ell$ are the vector and axial vector $Z\ell^+\ell^-$ couplings and are given as

$$C_V^\ell = -\frac{1}{2} + 2\sin^2\theta_W, \quad C_A^\ell = -\frac{1}{2}, \tag{29}$$

where $\theta_W$ is the weak mixing (or Weinberg) angle [44]. The contributions to decay $\Lambda_b \to \Lambda \ell^+ \ell^-$ mainly come from the Wilson coefficients $C_9$ and $C_{10}$, and corresponding operators. In this model, the structure of the effective Hamiltonian is in the same form as that of the SM. Hence, its effect can be evaluated by replacing the SM Wilson coefficients $\left(C_9^{eff}\right)^{SM}$ and $\left(C_{10}\right)^{SM}$ by

$$C_9^{eff} = \left(C_9^{eff}\right)^{SM} + \frac{2\pi}{\alpha} \frac{U_{sb}}{V_{tb} V_{ts}^*} ,$$

$$C_{10}^{eff} = \left(C_{10}\right)^{SM} - \frac{2\pi}{\alpha} \frac{U_{sb}}{V_{tb} V_{ts}^*} . \qquad (30)$$

From equation (2), it is clear that the value of the Wilson coefficients $C_9$ and $C_{10}$ are opposite to each other. Hence, the new physics contributions to $C_9$ and $C_{10}$ are opposite to each other and one will get constructive and destructive interference of SM and NP amplitudes for $\theta = \pi$ or zero (where $\theta$ is the relative weak phase between the SM and NP contribution in the above equation). We consider the weak phase difference to be $\pi$ to get constructive interference between the SM and NP amplitudes. The updated model-independent values for these coefficients can be found in Ref. [45,46].

The branching ratios for $\Lambda_b \to \Lambda \ell^+ \ell^-$ $(\ell = \mu, \tau)$ decay in the presence of Z-mediated FCNC are calculated in [2] as:

$$B\left(\Lambda_b \to \Lambda \mu^+ \mu^-\right)_Z = 20.90 \times 10^{-6} ,$$

$$B\left(\Lambda_b \to \Lambda \tau^+ \tau^-\right)_Z = 0.92 \times 10^{-6} . \qquad (31)$$

The same idea can be applied to a $Z'$-boson i.e., mixing among particles which have different $Z'$ quantum numbers will induce FCNCs due to $Z'$ exchange [31,32,47]. Since the $U_{pq}^Z$ are generated by mixing that breaks weak isospin, they are expected to be at most $O(M_1/M_2)$, where $M_1(M_2)$ is typical light (heavy) fermion mass. On the other hand, the $Z'$-mediated coupling $U_{pq}^{Z'}$ can be generated via mixing of particles with same weak isospin and, so, suffer no suppression. Even though $Z'$-mediated interactions are suppressed relative to Z, these are compensated by the factor $U_{pq}^{Z'} / U_{pq}^Z \sim (M_2/M_1)$. The flavor-changing coupling $U_{sb}^{Z'}$ is constrained by the process $b \to s \nu \bar{\nu}$ [48,49] and is found to be $\left|U_{sb}^{z'}\right| \frac{M_z^2}{M_{z'}^2} \leq 7.1 \times 10^{-3}$. This can be turned into a bound on $U_{sb}^{Z'}$ if one assumes a value for $M_{Z'}$. If we assume $\left|U_{sb}^{Z'}\right| \sim \left|V_{tb} V_{ts}^*\right|$, then it is possible to write $U_{sb}$ instead of $U_{sb}^{Z'}$, which gives significant contributions to the $\Lambda_b \to \Lambda \ell^+ \ell^-$ decay process. Thus the new contributions from $Z'$-boson are exactly in the similar manner as in the Z-boson. Therefore, we write the general effective Hamiltonian [31,32] that contribute to $\Lambda_b \to \Lambda \ell^+ \ell^-$ in the light of equation (28) as :

$$H_{eff}(Z') = \frac{G_F}{\sqrt{2}} U_{sb} \left[\bar{s}\gamma^\mu(1-\gamma_5)b\right]\left[\bar{\ell}\left(C_V^\ell \gamma_\mu - C_A^\ell \gamma_\mu\gamma_5\right)\ell\right]\left(\frac{g'}{g}\frac{M_Z}{M_{Z'}}\right)^2, \quad (32)$$

where $g = e/(\sin\theta_W \cos\theta_W)$ and $g'$ is the gauge coupling associated with the $U(1)'$ group. The absence of the suppression in the mixing for the $Z'$-mediated FCNC can compensate for $M_Z^2/M_{Z'}^2$ suppression of the $Z'$ amplitude relative to the Z amplitude, which implies that the coefficients describing the Z and $Z'$ flavor-changing effective interactions can be comparable in size. The net effective Hamiltonian can be written, from equation (28) and (32), as $H_{eff} = H_{eff}(Z) + H_{eff}(Z')$ and

$$H_{eff} = \frac{G_F}{\sqrt{2}} U_{sb} \left[\bar{s}\gamma^\mu(1-\gamma_5)b\right]\left[\bar{\ell}\left(C_V^\ell \gamma_\mu - C_A^\ell \gamma_\mu\gamma_5\right)\ell\right]\left[1 + \left(\frac{g'}{g}\frac{M_Z}{M_{Z'}}\right)^2\right], \quad (33)$$

and the corresponding branching ratios for the baryonic rare decays $\Lambda_b \to \Lambda \ell^+\ell^-$ ($\ell = \mu, \tau$) are calculated in the next section.

## 4. Results and Discussions

In this section, we calculate the branching ratios for the baryonic rare decays $\Lambda_b \to \Lambda \ell^+\ell^-$ ($\ell = \mu, \tau$) using recent data [41]: $m_\mu = (105.658367 \pm 0.000004)$ MeV, $m_\tau = (1776.84 \pm 0.17)$ MeV, $m_e = (0.510998910 \pm 0.000000013)$ MeV, $m_\Lambda = (1115.683 \pm 0.006)$ MeV, $m_{\Lambda_b} = (5620.2 \pm 1.6)$ MeV, $\tau_{\Lambda_b} = (1.383^{+0.049}_{-0.048}) \times 10^{-12}$ s, $M_Z = (91.1876 \pm 0.0021)$ GeV, $G_F = (1.16637 \pm 0.00001) \times 10^{-5}$ GeV$^{-2}$, $\sin^2\theta_W = 0.23$ and $|U_{sb}| \cong 10^{-3}$ [42]. Since the $Z'$ has not yet been discovered, its mass is unknown. However, the $Z'$ mass is constrained by direct searches at Fermilab, weak neutral current data and precision studies at LEP and the SLC [50-53], which give a model-dependent lower bound around 500 GeV if the interaction is comparable to the other couplings of the standard model. However, the lower mass limit can be as low as 130 GeV [54] if the coupling is weak. The experimental bounds on $M_{Z'}$ above 1 TeV are not expected without excessive fine-tuning of supersymmetry breaking mass parameters, or unusual choices of $U(1)'$ charge assignments. In a study of $B$ meson decays with $Z'$-mediated flavor-changing neutral currents [32], they study the $Z'$-boson in the mass range of a few hundred GeV to 1 TeV. In this paper, we study the $Z'$-boson in the mass range 130 GeV – 1 TeV.

In general, the value of $g'/g$ is undetermined [55]. However, generically, one expects that $g'/g \approx 1$ if both U(1) groups have the same origin from some grand unified theory. We take $g'/g \approx 1$ in our calculations.

Using the lower limit for the mass of Z′-boson, $M_{Z'}$ = 130 GeV, we get

$$B(\Lambda_b \to \Lambda\mu^+\mu^-)_{Z+Z'} = (46.52 \pm 0.01) \times 10^{-6},$$
$$B(\Lambda_b \to \Lambda\tau^+\tau^-)_{Z+Z'} = (2.03 \pm 0.12) \times 10^{-6}. \tag{34}$$

Again using the mass of Z′-boson, $M_{Z'}$ = 1000 GeV, we get

$$B(\Lambda_b \to \Lambda\mu^+\mu^-)_{Z+Z'} = (21.23 \pm 0.11) \times 10^{-6},$$
$$B(\Lambda_b \to \Lambda\tau^+\tau^-)_{Z+Z'} = (0.93 \pm 0.02) \times 10^{-6}. \tag{35}$$

From equation (34) and (35), it is clear that depending on the precise value of $M_{Z'}$, the Z′-mediated FCNCs gives sizable contributions to $\Lambda_b \to \Lambda\ell^+\ell^-$ decay process. Our estimated branching ratios for $\Lambda_b \to \Lambda\ell^+\ell^-$ decay process are reasonably enhanced from its standard model value [equation (20)]. Hence, the $\Lambda_b \to \Lambda\ell^+\ell^-$ decay process could provide signals for new physics beyond the standard model. It is also found that the forward-backward asymmetries ($A_{FB}$) are different from that of the standard model value due to Z-mediated FCNC [2]. We expect that the Z′-mediated FCNC will also change the forward-backward asymmetries values from that of the standard model value. The position of the zero value of $A_{FB}$ is very sensitive to the presence of new physics. These facts lead to enrichment in the phenomenology of both the Z and Z′-mediated FCNCs and $\Lambda_b \to \Lambda\ell^+\ell^-$ decay; and the physics beyond the standard model will be known after the discovery of the Z′-boson which is expected at the LHC.

We are hopeful that the effect of Z′-mediated FCNCs in $\Lambda_b$ decays would be measured in the Tevatron. The LHCb is expected to give us more insight on the flavor structure of new physics through precise measurements of rates and CP asymmetries in rare decays. The experimental observation of the rare B-decays $\Lambda_b \to \Lambda\ell^+\ell^-$ would provide precision tests of the SM in the crucial and as yet untested FCNC sector of B-decays. It is also expected [18,22,56-58] that the measurements of $\Lambda_b \to \Lambda\ell^+\ell^-$ can serve as a promising quantity to explore new physics effects as well as to constrain the parameter space of various models beyond the SM. We have no doubt that an exciting future is ahead of us!

**Acknowledgments**

We gratefully acknowledge helpful and enlightening discussions with Prof. T. M. Aliev, Middle East Technical University, Turkey; Dr. A. K. Giri, Punjabi University, Patiala and Dr. R. Mohanta, University of Hyderabad, India. We also thank Dr. Kartik Senapati, Cambridge University and Dr. S. C. Martha & Mr. Sudhansu Biswal, IISc. Bangalore for their help in the preparation of the manuscript. We thank the referee for suggesting valuable improvements of our manuscript.

**References**

1. Mohanta R 2005 *Phys. Rev.* D **71** 114013 [hep-ph/0503225]
2. Giri A K and Mohanta R 2006 *Eur. Phys. J.* C **45** 151
3. Rou W S, Willey R S and Soni A 1987 *Phys. Rev. Lett.* **58** 1608
4. Deshpande N G and Trampetic J 1988 *Phys. Rev. Lett.* **60** 2583
5. Lim C S, Morozumi T and Sanda A I 1989 *Phys. Lett.* B **218** 343
6. Grinstein B, Savage M J and Wise M B 1989 *Nucl. Phys.* B **319** 271
7. Dominguez C, Paver N and Riazuddin 1988 *Phys. Lett.* B **214** 459
   Paver N and Riazuddin 1992 *Phys. Rev.* D **45** 978
8. Deshpande N G, Trampetic J and Ponose K 1989 *Phys. Rev.* D **39** 1461
9. Ali A, Mannel T and Morozumi T 1991 *Phys. Lett.* B **273** 505
10. Ali A, Giudice G F and Mannel T 1995 *Z. Phys.* C **67** 417
11. Greub C, Ioannissian A and Wyler D 1995 *Phys. Lett.* B **346** 149
    Liu D 1995 *Phys. Lett.* B **346** 355
    Burdman G 1995 *Phys. Rev.* D **52** 6400
    Okada Y, Shimizu Y and Tanaka M 1997 *Phys. Lett.* B **405** 297
12. Deshpande N G, He X – G and Trampetic J 1996 *Phys. Lett.* B **367** 362
13. Bertolini S, Borzumati F, Masiero A and Ridolfi G 1991 *Nucl. Phys.* B **353** 591
14. Krüger F and Sehgal L M 1997 *Phys. Rev.* D **55** 2799
    Krüger F and Sehgal L M 1997 *Phys. Rev.* D **56** 5452
15. Huang C-S and Yan H-J 1999 *Phys. Rev.* D **59** 114022
    Huang C-S and Yan H-J 2000 *Phys. Rev.* D **61** 039901 (Erratum)
16. Chen C-H and Geng C Q 2001 *Phys. Lett.* B **516** 327 [hep-ph/0101201]
17. Chen C-H and Geng C Q 2001 *Phys. Rev.* D **63** 114024 [hep-ph/0101171]
18. Chen C-H and Geng C Q 2001 *Phys. Rev.* D **64** 074001 [hep-ph/0106193]
19. Chen C-H, Geng C Q and Ng J N 2000 *Phys. Rev.* D **65** 091502
    Chen C-H, Geng C Q and Ng J N 2001 arXiv:hep-ph/0210067.
20. Aliev T M and Savci M 2000 *J. Phys.* G **26** 997
21. Aliev T M, Ozpineci A and Savci M 2003 *Nucl. Phys.* B **649** 168 [hep-ph/ 0202120]
    Aliev T M, Ozpineci A and Savci M 2002 *Phys. Rev.* D **65** 115002 [hep-ph/0203045]
    Aliev T M, Ozpineci A and Savci M 2003 *Phys. Rev.* D **67** 035007 [hep-ph/0211447]
    Aliev T M, Ozpineci A and Savci M 2005 *Nucl. Phys.* B **709** 115 [hep-ph/0407217]
    Aliev T M, Ozpineci A, Savci M and Yuce C 2002 *Phys. Lett.* B **542** 229
    [hep-ph/0206014].
22. Wang Y-M, Aslam M J and Lu C-D 2009 *Eur. Phys. J.* C **59** 847
23. Langacker P 2003 arXiv: hep-ph/0308033
    Langacker P 2008 arXiv:0801.1345 [hep-ph]
24. Fuks B 2008 arXiv:0805.2004 [hep-ph]
25. Leike A 1999 *Phys. Rep.* **317** 143 [arXiv:hep-ph/9805494]
26. Hewett J and Rizzo T 1989 *Phys. Rep.* **183** 193
27. Sahoo S 2008 *Indian J. Phys.* **80** (2) 191
    He X-Gand Valencia G 2006 arXiv:hep-ph/0605202
    Feldman D, Liu Z and Nath P 2006 *Phys. Rev. Lett.* **97** 021801 [hep-ph/0603039]
    Deo B B and Maharana L 1999 *Phys. Lett.* B **461** 105
28. Abe F *et al.* [CDF Collaboration] 1996 *Phys. Rev. Lett.* **77** 438
    The LEP Collaboration and LEP Electroweak Working Group, CERN-PPE / 95 / 172.

29. Babu K S, Kolda C and March-Russell J 1996 *Phys. Rev.* D **54** 4635 [hep-ph/9603212]
    Babu K S, Kolda C and March-Russell J 1998 *Phys. Rev.* D **57** 6788 [hep-ph/9710441]
30. Rizzo T G 1998 *Phys. Rev.* D **59** 015020 [hep-ph/9806397]
    del Aguila F, Moreno J and Quiros M 1991 *Nucl. Phys.* B **372** 3
    del Aguila F, Moreno J and Quiros M 1991 *Nucl. Phys.* B **361** 45
31. Sahoo S and Maharana L 2004 *Phys. Rev.* D **69** 115012
    Barger V, Chiang C-W, Langacker P and Lee H S 2004 *Phys. Lett.* B **598** 218
    Barger V, Chiang C-W, Jiang J and Langacker P 2004 *Phys. Lett.* B **596** 229
32. Barger V, Chiang C-W, Langacker P and Lee H S 2004 *Phys. Lett.* B **580** 186 [hep-ph/0310073]
33. Arhrib A, Chung K, Chiang C-W and Yuan T-C 2006 arXiv:hep-ph/0602175
34. Cheung K, Chiang C-W, Deshpande N G and Jiang J 2006 arXiv:hep-ph/0604223
35. Nir Y and Silverman D 1990 *Phys. Rev.* D **42** 1477
    Silverman D 1992 *Phys. Rev.* D **45** 1800
    Barger V, Berger M S and Phillips R J N 1995 *Phys. Rev.* D **52** 166
    Gronau M and London D 1997 *Phys. Rev.* D **55** 2845
36. Buchalla G, Buras A J and Lautenbacher M E 1996 *Rev. Mod. Phys.* **68** 1125 [arXiv:hep-ph/9512380]
    A. Ali 1996 arXiv:hep-ph/ 9606324
37. Beneke M, Feldmann Th. and Seidel D 2001 *Nucl. Phys.* B **612** 25 [arXiv:hep-ph/0106067]
38. Buras A J and M*ü*nz M 1995 *Phys. Rev.* D **52** 186
39. O'Donnell P J, Sutherland M and Tung H K 1992 *Phys. Rev.* D **46** 4091
    Kr*ü*ger F and Sehgal L M 1996 *Phys. Lett.* B **380** 199
40. Mannel T, Roberts W and Ryzak Z 1991 *Nucl. Phys*. B **355** 38
41. Amsler C *et al.* [Particle Data Group] 2008 *Phys. Lett.* B **667** 1
42. Giri A K and Mohanta R 2003 *Phys. Rev.* D **68** 014020 [arXiv:hep-ph/0306041]
43. Kaneko J *et al.* [BELLE Collaboration] 2003 *Phys. Rev. Lett.* **90** 021801
44. Glashow S L 1961 *Nucl Phys*. **22** 579
    Weinberg S 1967 *Phys. Rev. Lett.* **19** 1264
    Salam A 1968 in *Elementary particle physics:* Nobel Symp. No. 8, Ed. N. Svartholm, Stockholm.
45. Altmannshofer W *et al*. 2009 *JHEP*. **0901** 019 [arXiv:0811.1214 (hep-ph)].
46. Hurth T, Isidori G, Kamenik J F and Mescia F 2009 *Nucl. Phys.* B **808** 326 [arXiv: 0807.5039 (hep-ph)].
47. Nardi E 1993 *Phys. Rev.* D **48** 1240
    Bernabeu J, Nardi E and Tommasini D 1993 *Nucl. Phys.* B **409** 69 [hep-ph/9306251]
48. Grossman Y, Ligeti Z and Nardi E 1996 *Nucl. Phys.* B **465** 369
    Grossman Y, Ligeti Z and Nardi E 1996 *Nucl. Phys.* B **480** 753
49. Leroux K and London D 2002 *Phys. Lett.* B **526** 97
50. Abe F *et al.* [CDF Collaboration] 1991 *Phys. Rev. Lett.* **67** 2418
    Abe F *et al.* [CDF Collaboration] 1997 *Phys. Rev*. **79** 2192
51. Abulencia A *et al.* [CDF Collaboration] 2006 *Phys. Rev. Lett.* **96** 211801

52. The LEP Electroweak Working Group and SLD Heavy Flavour Group 2002 arXiv:hep-ex/0212036.
53. For review, see Cvetic M and Langaker P 1997 arXiv:hep-ph/9707451
54. Cvetic M and Langaker P 1996 *Mod. Phys. Lett.* A **11** 1247 [arXiv:hep-ph/9602424]
55. Cvetic M and Lynn B W 1987 *Phys. Rev.* D **35** 51
56. Wang Y-M, Li Y and Lu C-D 2009 *Eur. Phys. J.* C **59** 861 [arXiv: 0804.0648 (hep-ph)].
57. Hiller G and Kagan A 2002 *Phys. Rev.* D **65** 074038 [arXiv:hep-ph/0108074 v2 29 Jan 2002]
58. Mannel T and Recksiegel S 1998 *J. Phys.* G **24** 979